\documentclass[doublecol]{epl2}
\usepackage{graphicx}
\usepackage{fleqn}
\usepackage{amsmath,amssymb,revsymb,epsfig}
\usepackage{color}

\title{Reverse engineering of complex dynamical networks in the presence of time-delayed interactions based on noisy time series}

\author{Wen-Xu Wang\inst{1}, Jie Ren\inst{2,3,5}, Ying-Cheng Lai\inst{1,3,4},
Baowen Li\inst{2,3}}

\institute{ \inst{1} School of Electrical, Computer and Energy
Engineering, Arizona State University, Tempe, AZ 85287  \\
\inst{2} NUS Graduate School for Integrative Sciences and
Engineering, Singapore 117456, Republic of Singapore \\
\inst{3} Department of Physics and Centre for Computational
Science and Engineering, National University of Singapore,
Singapore 117546, Republic of Singapore\\
\inst{4} Department of Physics, Arizona State University, Tempe,
Arizona 85287, USA\\
\inst{5} Theoretical Division, Los Alamos National Laboratory, Los Alamos, New Mexico 87545, USA}

\abstract{Reverse engineering of complex dynamical networks is important for a variety of fields where uncovering the full topology of unknown networks and estimating parameters characterizing the network structure and dynamical processes are of interest. We consider complex oscillator networks with time-delayed interactions in a noisy environment, and develop an effective method to infer the full topology of the network and evaluate the amount of time delay based solely on noise- contaminated time series. In particular, we develop an analytic theory establishing that the dynamical correlation matrix, which can be constructed purely from time series, can be manipulated to yield both the network topology and the amount of time delay simultaneously. Extensive numerical support is provided to validate the method. While our method provides a viable solution to the network inverse problem, significant difficulties, limitations, and challenges still remain, and these are discussed thoroughly.}

\begin{document}
\maketitle

\textbf{Time-delayed interactions are common in complex systems arising from various fields of science and engineering. Con- sider, for example, a coupled oscillator network in a physi- cal environment where noise is present. Time delay can typically occur in the node-to-node interactions. Now sup- pose that no prior knowledge about the nodal dynamics and the network topology is available but only a set of noise- contaminated time series can be obtained through measure- ments. The question is whether it is possible to deduce the full topology of the network and to estimate the amount of average time delay using the time series only. This issue belongs to the recently emerged subfield of research in com- plex systems: reverse engineering of complex networks (or the inverse problem). While a number of methods address the network inverse problem have appeared, to our knowl- edge, the issue of time-delayed interactions has not been considered. Here we present an effective method to infer the full network topology and, at the same time, to estimate the amount of average time delay in the network. In partic- ular, we develop a physical theory to obtain a formula relat- ing the network topology and time delay to the dynamical correlation matrix, which can be constructed purely from time series. We then show how information about the time delay encrypted in the dynamical correlation matrix can be separated from that of network topology, allowing both to be inferred in a computationally extremely efficient man- ner. We present numerical examples from both model and real-world complex networks to demonstrate the working of our method. Difficulties, limitations, and challenges are also discussed. Reverse-engineering of complex dynamical systems has potential applications in many disciplines, and our work represents a small step forward in this extremely challenging area.}

One of the outstanding issues in nonlinear and statistical physics,
and also in network science and engineering, is to infer or predict
the topology and other basic characteristics of complex networks
based only on measured time series. This ``inverse'' problem is
relevant to a number of fields such as biomedical and techno-social
sciences where complex networked systems are ubiquitous. In defense,
the problem of identifying various adversarial networks based on
observations is also of paramount importance. Despite tremendous
efforts in revealing the connection between network structures and
dynamics \cite{PC:1998,dyn1,dyn2,dyn3,dyn4}, how to infer the
underlying topology from dynamical behaviors is still challenging as
an inverse problem, especially in the absence of the knowledge of
nodal dynamics. 

Recent years have witnessed the emergence of a number of methods 
to address various aspects of the inverse problem, which include gene networks
inference using singular value decomposition and robust regression
\cite{collins}, spike classification methods for measuring
interactions among neurons from spike trains \cite{spike},
symbolically reverse engineering of coupled ordinary differential
equations \cite{Lipson}, approaches based on response dynamics of a
specific oscillators \cite{Timme:2007}, $L_1$ norm in optimization
theory \cite{NS:2008}, noise induced scaling laws \cite{WCHLH:2009},
and the interplay between dynamical correlation and network
structure in the presence of noise \cite{RWLL:2010}. However, the
issue of time delay has not been addressed yet. The purpose of this
article is to present a general theory that leads to a completely
data-driven and extremely efficient method to predict the network
topology and the time delay at the same time.

Time delay is fundamental in natural systems, due to the finite
propagation speed of physical signals. In addition to numerous
examples in physics, situations where time delay is important
include the latency times of neuronal excitations in neuroscience,
finite reaction times of chemicals in chemistry, etc.. In coupled
oscillator networks, the effects of time delay on dynamics under
various {\em given} network topologies have been studied extensively
\cite{Ding:2004,Kanter:2009,Yan:2009,Scholl:2010,HKS:2010}. In our
case, however, the network connections, the amount of the time
delay, and other properties of the network are unknown {\it a
priori}, and our goal is to predict these by using noisy time series
only. To be concrete, we shall focus on complex oscillator networks.
Our general point of view is that, information about the network
topology and time delay has been encoded in noisy time series from
various nodes in the network. The objective of solving the inverse
problem is to decode such information from noisy time series.

Our idea is that, if the networked system suffers from a noisy
environment so that the measured time series are noisy, it is
possible to accomplish the task of decoding in a natural way. In
particular, we construct a dynamical correlation matrix from all
available time series, the elements of which are the average
products of the deviations of all pair-wise time series from a mean
value. We shall show analytically that information about the network
structure and time delay can be decoded through this matrix. In
fact, as we will show in developing our theory, information about
the network topology can be separated from the time delay through a
generalized inverse operation of the dynamical correlation matrix,
enabling a complete prediction of the underlying networked system.

To provide numerical support for our theory, we exploit three
representative dynamical systems on homogeneous and heterogeneous
model complex networks and on a number of real-world networks as
well. We find that the presence of a time delay results in a
deviation in the distribution of the diagonal elements of the
dynamical correlation matrix from a power law, which can be used as
a preliminary criterion to determine whether there is a significant
time delay in the underlying networked system. Computations reveal
high accuracies in the prediction of both the network topology and
the time delay for all combinations of dynamical systems and network
models studied.

We present our theory and method by considering a network of $N$
coupled oscillators. Each oscillator, when decoupled, satisfies
$\dot{\vec{\mathrm{x}}}_i = \mathbf{F}_i[\vec{\mathrm{x}}_i]$, where
$\vec{\mathrm{x}}_i$ denotes the $d$-dimensional state variable of
node $i$. The dynamics of the whole time-delayed system in a noisy
environment is described as:
\begin{equation} \label{eq:main}
\dot{\vec{\mathrm{x}}}_i(t) = \mathbf{F}_i[\vec{\mathrm{x}}_i(t)] -
c \sum^N_{j=1} L_{ij} \mathbf{H}[\vec{\mathrm{x}}_j(t-\tau)] +
\vec{\eta}_i(t),
\end{equation}
where $c$ is the coupling strength and $\mathbf{H}$ denotes the
coupling function. $L_{ij}$ is Laplacian matrix, characterizing the
topology of the underlying network that $L_{ij}=-1$ if $j$ connects
to $i$ (otherwise $0$) for $i\neq j$, and $L_{ii}=k_i$, where $k_i$
is the degree of node $i$. $\tau$ denotes the time delay, and
$\vec{\eta}_i$ is a $d$-dimensional stochastic process representing
noise on node $i$ (In the following, we use $\;\vec{}\;$ on the head
to denote the $d$-dimensional state variable). The standard
procedure of linearization \cite{Ding:2004,PC:1998} can be carried
out by letting $\vec{\mathrm{x}}_i = \bar{\mathrm{x}}_i +
\vec{\xi}_i$, where $\mathrm{\bar{x}}_i$ is the counterpart of
$\mathrm{x}_i$ in the absence of noise. The $d$-dimensional
dynamical process governing the fluctuations on $i$th oscillator can
then be obtained as the variational equation:
\begin{equation} \label{eq:linearize}
\dot{\vec{\xi}}_i(t) = D\mathbf{F}_i\cdot\vec{\xi}_i(t) -
c\sum_{j=1}^{N}L_{ij} D\mathbf{H}\cdot\vec{\xi}_j(t-\tau) +
\vec{\eta}_i(t),
\end{equation}
where $D\mathbf{F}_i$ and $D\mathbf{H}$ denote the $d\times d$
Jacobian matrices of the intrinsic dynamics $\mathbf{F}_i$ and the
coupling function $\mathbf{H}$, respectively. Decomposing
Eq.~(\ref{eq:linearize}) in terms of the eigenmodes, we obtain
\begin{equation}
\dot{\vec{\epsilon}}_{\alpha}(t) = \sum_{\beta} D\mathbb{F}_{\alpha
\beta}\cdot \vec{\epsilon}_{\beta}(t) -
c\lambda_{\alpha}D\mathbf{H}\cdot\vec{\epsilon}_{\alpha}(t-\tau) +
\vec{\zeta}_{\alpha}(t).
\end{equation}
Here, instead of the index $i,j$ running on the real space of
networks, the index $\alpha, \beta$ run on the eigen-space.
$\vec{\epsilon}_{\alpha}=\sum_i\psi_{\alpha i}\vec{\xi}_i$,
$\vec{\zeta}_{\alpha}=\sum_i \psi_{\alpha i}\vec{\eta}_i$ and
$D\mathbb{F}_{\alpha \beta}= \sum_{i}\psi_{\alpha
i}D\mathbf{F}_i\psi_{\beta i}$, where $\psi_{\alpha j}$ denotes the
$\alpha$th normalized eigenvector of the Laplacian matrix, and
$\lambda_{\alpha}$ is the corresponding eigenvalues that satisfy
$0=\lambda_{0}<\lambda_{1}\leq\cdots\leq\lambda_{N-1}$. Under the
approximation $D\mathbf{F}_i \approx D\mathbf{F}$ so that
$D\mathbb{F}_{\alpha \beta} = D\mathbf{F}\delta_{\alpha \beta}$, the
above equation can be reduced to
\begin{equation}
\dot{\vec{\epsilon}}_{\alpha}(t)=D\mathbf{F}\cdot\vec{\epsilon}_{\alpha}(t)
-c\lambda_{\alpha}D\mathbf{H}\cdot\vec{\epsilon}_{\alpha}(t-\tau) +
\vec{\zeta}_{\alpha}(t).
\end{equation}
From the covariance of Gaussian noise $\langle
\vec{\eta}_i(t)\vec{\eta}_j^T(t') \rangle = \sigma^2\mathbf{I}_d
\delta_{ij}\delta (t-t')$ with $\mathbf{I}_d$ the $d$-dimensional
identity matrix and $\sigma^2$ the noise strength, we obtain
$\langle \vec{\zeta}_{\alpha}(t)\vec{\zeta}_{\beta}^T(t') \rangle =
\sigma^2 \mathbf{I}_d\delta_{\alpha \beta}\delta (t-t')$, which
indicates the stochastic process we mapped into eign-space is still
Gaussian noise. Assuming small time delay, we can apply the
first-order approximation: $\vec{\epsilon}_{\alpha}(t-\tau) =
\vec{\epsilon}_{\alpha}(t) - \tau \dot{\vec{\epsilon}}_{\alpha}(t)$,
which yields
$$(\mathbf{I}_d-c\tau\lambda_{\alpha}D\mathbf{H})\dot{\vec{\epsilon}}_{\alpha}(t)
= -(c\lambda_{\alpha}D\mathbf{H}-D\mathbf{F}
)\vec{\epsilon}_{\alpha}(t) + \vec{\zeta}_{\alpha}(t).$$
Denote
$\mathbf{B}=(\mathbf{I}_d-c\tau\lambda_{\alpha}D\mathbf{H})^{-1}$,
$\mathbf{A}=\mathbf{B}(c\lambda_{\alpha}D\mathbf{H}-D\mathbf{F})$,
and follow the standard stochastic calculus \cite{Gardiner}, we get
the solution:
\begin{eqnarray}
\vec{\epsilon}_{\alpha}(t)=e^{-\mathbf{A}t}\vec{\epsilon}_{\alpha}(0)
+\int^t_0e^{-\mathbf{A}(t-t')}\mathbf{B}\vec{\zeta}_{\alpha}(t')dt'.
\end{eqnarray}
Since we are interested in the regime where oscillator states are
perturbed from the synchronized manifold by the noisy environment,
we assume the system is in the absence of divergence of state
variables, therefore in the long time limit, the initial condition
term can be discarded and we have \cite{RWLL:2010}:
$$\mathbf{A}\langle\vec{\epsilon}_{\alpha}\vec{\epsilon}_{\alpha}^T\rangle+
\langle\vec{\epsilon}_{\alpha}\vec{\epsilon}_{\alpha}^T\rangle\mathbf{A}=
\sigma^2\mathbf{B}\mathbf{B}^T.$$
The general solution of
$\langle\vec{\epsilon}_{\alpha}\vec{\epsilon}_{\alpha}^T\rangle$,
the $d\times d$ covariance matrix about $d$-dimensional states of
the $\alpha$th oscillator in the eigen-space, can be written as
\cite{Horn}:
$${vec}(\langle\vec{\epsilon}_{\alpha}\vec{\epsilon}_{\alpha}^T\rangle)
=\sigma^2{vec}(\mathbf{B}\mathbf{B}^T)/(\mathbf{I}_d\otimes\mathbf{A}+
\mathbf{A}\otimes\mathbf{I}_d),$$ where the operator
$vec(\mathbf{X})$ creates a column vector from a matrix $\mathbf{X}$
by stacking the columns of $\mathbf{X}$ below one another.

Although we obtain this solution, it is not practical in real
applications. In follows, we approximate the state variables as one
dimension such that $D\mathbf{H}=1$ and drop the notation $\vec{}$ .
In this way, the above solution is simplified to:
\begin{eqnarray}
\langle \epsilon_{\alpha}^2 \rangle =
\frac{\sigma^2}{2c}\frac{1}{(1-c\tau
\lambda_{\alpha})(\lambda_{\alpha} - {D\mathbf{F}}/{c})}.
\end{eqnarray}
Return to real variables from the eigen-space by inserting $\xi_i
=\sum_{\alpha} \psi_{\alpha i}\epsilon_{\alpha}$ into the
correlation function $C_{ij} = \langle \xi_i \xi_j\rangle$ in the
real space between any two nodes, we have $C_{ij} = \sum_{\alpha =
1}^{N-1} \psi_{\alpha i} \psi_{\alpha j} \langle \epsilon_{\alpha}^2
\rangle $ such that
\begin{eqnarray} \label{eq:return}
C_{ij} = \frac{\sigma^2}{2c}\sum_{\alpha =1}^{N-1}
\frac{\psi_{\alpha i}\psi_{\alpha j}}{( 1- c\tau \lambda_{\alpha}
)(\lambda_{\alpha} - {D\mathbf{F}}/{c}) }.
\end{eqnarray}
Under the approximation of negligible $D\mathbf{F}/c$ and reminding
of the small time delay $\tau$, Eq. (\ref{eq:return}) for the
dynamical correlation can then be expanded as:
\begin{equation} \label{eq:forL}
C_{ij} \approx \frac{\sigma^2}{2c} \sum_{\alpha
=1}^{N-1}\frac{1+c\tau
\lambda_{\alpha}}{\lambda_{\alpha}}\psi_{\alpha i}\psi_{\alpha j} =
\frac{\sigma^2}{2c}[\mathbf{L}^\dag + c\tau \mathbf{I}_N]_{ij},
\end{equation}
wherein under the effect of noise $\sigma^2$, the dynamics in terms
of correlation matrix $\mathbf{{C}}$ is connected explicitly with
the time delay $\tau$ and the structure information in terms of
$\mathbf{{L}^\dag}=\sum^{N-1}_{\alpha=1}{\psi_{\alpha i}\psi_{\alpha
j}}/{\lambda_{\alpha}}$, the pseudo-inverse of Laplacian matrix.
Note that the time delay has no effect on the cross-correlation
elements except the auto-correlations due to the identity matrix.
Following Eq.~(\ref{eq:forL}), the diagonal elements $C_{ii}$ of the
dynamical correlation matrix can be obtained by expanding
$\mathbf{{L}}^{\dag}$ in terms of the underlying network structure
\cite{RWLL:2010}:
\begin{eqnarray} \label{eq:cii}
C_{ii} &\approx& \frac{\sigma^2}{2c}[\mathbf{{K}}^{-1} +
\mathbf{{K}}^{-1}\mathbf{{P}}\mathbf{{K}}^{-1} +
\mathbf{{K}}^{-1}\mathbf{{P}}\mathbf{{K}}^{-1}\mathbf{{P}}\mathbf{{K}}^{-1}]_{ii}
\nonumber\\ &+& \frac{\sigma^2 \tau}{2} \approx \ \
\frac{\sigma^2}{2ck_i}\bigg(1+ \frac{1}{\langle k\rangle} \bigg) +
\frac{\sigma^2 \tau}{2},
\end{eqnarray}
where $\mathbf{{K}}=\mathrm{diag}(k_1,\cdots,k_N)$ is the degree
matrix, $\mathbf{{P}}$ is the adjacency matrix such that
$\mathbf{{L}}=\mathbf{{K}}-\mathbf{{P}}$ and $\langle k\rangle$
denotes the average degree. We see that the fluctuations $C_{ii}$ at
node $i$ depend both on its local structure $k_i$ and the time delay
$\tau$. When $\tau=0$, this result is consistent with the recently
discovered noise-induced scaling law \cite{WCHLH:2009}, derived
there by a power-spectral analysis.

The off-diagonal elements of $\mathbf{{L}}$ contain complete
information about the network structure while its diagonal elements
can be obtained from the off-diagonal ones. We thus focus on the
off-diagonal elements. For $i\neq j$, following
Eq.~(\ref{eq:return}), the generalized inverse matrix
$\mathbf{{C}}^{\dag}$ is
\begin{equation} \label{eq:fortau}
C_{ij}^{\dag} \approx
\frac{2c}{\sigma^2}\sum_{\alpha=1}^{N-1}\lambda_{\alpha}(1 - c\tau
\lambda_{\alpha})\psi_{\alpha i}\psi_{\alpha j}
 = \frac{2c}{\sigma^2}[\mathbf{{L}}- c\tau
\mathbf{{L}}^2]_{ij}.
\end{equation}
Considering $\mathbf{{L}}=\mathbf{{K}}-\mathbf{{P}}$, we can cast
this equation in the following form:
\begin{eqnarray}
C_{ij}^\dag=\frac{2c}{\sigma^2}[\mathbf{{L}}+c\tau(\mathbf{{K}}
\mathbf{{P}}+\mathbf{{P}}\mathbf{{K}}-\mathbf{{K}}^2-\mathbf{{P}}^2)]_{ij}.
\end{eqnarray}
For those off-diagonal elements $(i, j)$, the diagonal matrix
$\mathbf{K}^2$ has no contributions and $\mathbf{P}^2$ contributes
$l_{ij}$, where $l_{ij}$ is the number of two-step paths connecting
$i$ with $j$. By considering the negligible contribution of $l_{ij}$
compared with degrees, we thus have
\begin{eqnarray} \label{eq:infer}
\frac{\sigma^2}{2c}C_{ij}^\dag\approx \bigg\{
\begin{array}{cc}
  L_{ij}+c\tau(k_i+k_j), & \text{if
$i$ connects with $j$} \\
  0. & \text{otherwise} \\
\end{array}
\end{eqnarray}
Equation (\ref{eq:infer}) is one of our main results for network
inference, which indicates that the network structure can be
inferred through the off-diagonal elements $C_{ij}^{\dag}$ of the
dynamical correlation matrix based solely on the measured time
series.

Once $\mathbf{{L}}$ is predicted, the time delay $\tau$ can be
estimated, e.g., from Eq.~(\ref{eq:fortau}). We obtain
\begin{equation}
\tau \approx \bigg\langle \frac{\big[\mathbf{{L}}-
\frac{\sigma^2}{2c}\mathbf{{C}}^{\dag}\big]_{ij}}{c[\mathbf{{L}}^2]_{ij}}
\bigg\rangle _{i\neq j, L_{ij}\neq 0, (\mathbf{{L}}^2)_{ij} \neq 0},
\label{eq:tau}
\end{equation}
where the subscript in the average $\langle \cdot \rangle$ covers
all possible pairs of $i$ and $j$ by excluding the diagonal elements
in the matrices $\mathbf{{L}}$ and $\mathbf{{L}}^2$, and all pairs
with zero elements in the matrix $\mathbf{{L}}$ or $\mathbf{{L}}^2$.
Excluding zero elements can effectively reduce the estimation error
for $\tau$.

\begin{figure}
\begin{center}
\epsfig{file=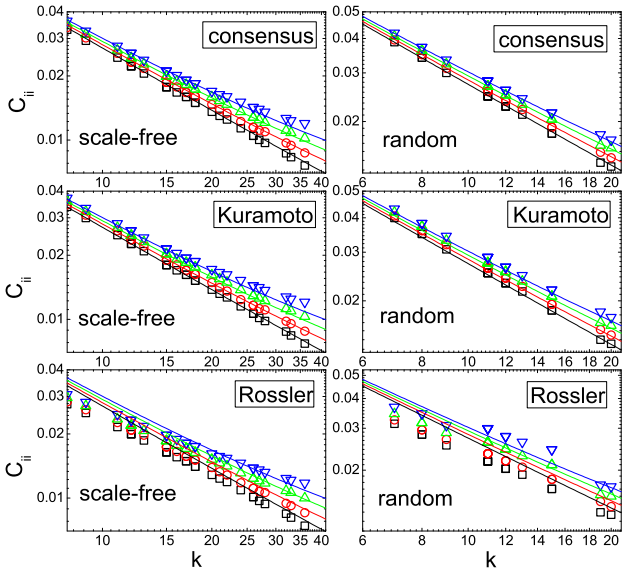,width=0.95\linewidth} \caption{(Color
online) Diagonal elements $C_{ii}$ of the dynamical correlation
matrix as a function of node degree $k$ for three dynamical
processes with different time delay $\tau$ on scale-free and random
networks. Square, circle, triangle and reverse triangle denote
$\tau=0.01$, 0.05, 0.07 and 0.09, respectively. The curves are the
theoretical prediction from Eq.~(\ref{eq:cii}).  The sizes of model
networks are 100 and the average degree is 10. The noise strength
$\sigma^2$ is 0.1 and the coupling strength $c$ is 0.2.
\label{fig:cii}}
\end{center}
\end{figure}

We now demonstrate numerically our method by considering several
model and real-world networks in the presence of noise and time
delay. For each network, we implement three dynamical processes: (i)
Consensus dynamics \cite{Saber:2007}:
$$\dot{x}_i(t) =
c\sum_{j=1}^{N}P_{ij}[x_j(t-\tau) - x_i(t-\tau)] +\eta_i,$$ (ii)
R\"ossler dynamics \cite{Rossler:1976}:
\begin{eqnarray}
\left\{%
\begin{array}{l}
    \dot{x}_i(t)=-y_i-z_i+c\sum_{j=1}^{N}P_{ij}[x_j(t-\tau) - x_i(t-\tau)]+\eta_i, \\
    \dot{y}_i = x_i + 0.2y_i + c\sum_{j=1}^{N} P_{ij}(y_j - y_i), \\
    \dot{z}_i = 0.2 + z_i(x_i - 9.0) + c\sum_{j=1}^{N}P_{ij}(z_j-z_i), \\
\end{array}%
\right. \nonumber
\end{eqnarray}
and (iii) Kuramoto phase oscillators \cite{Kuramoto:book}:
$$\dot{\theta}(t) = \omega_i + c\sum_{j=1}^{N} \sin [\theta_j(t-\tau)
- \theta_i(t-\tau)] + \eta_i,$$ where $\theta_i$ and $\omega_i$ are
the phase and the natural frequency of oscillator $i$.

Time series are then collected from all nodes. The element of the
dynamical correlation matrix between two arbitrary nodes $i$ and $j$
is calculated as $C_{ij}=\langle[x_i(t) - \bar{x}(t)]\cdot [x_j(t) -
\bar{x}(t)] \rangle_t$, where $\bar{x}(t) = (1/N)\sum_{i=1}^{N}
x_i(t)$ and $\langle \cdot \rangle_t$ denotes the long time average.
For the R\"ossler dynamics, $x_i(t)$ stands for the $x$ component of
the $i$th oscillator and for the Kuramoto dynamics, $x_i(t)$ stands
for the phase variable $\theta_i(t)$ of the $i$th oscillator.

\begin{figure}
\begin{center}
\epsfig{file=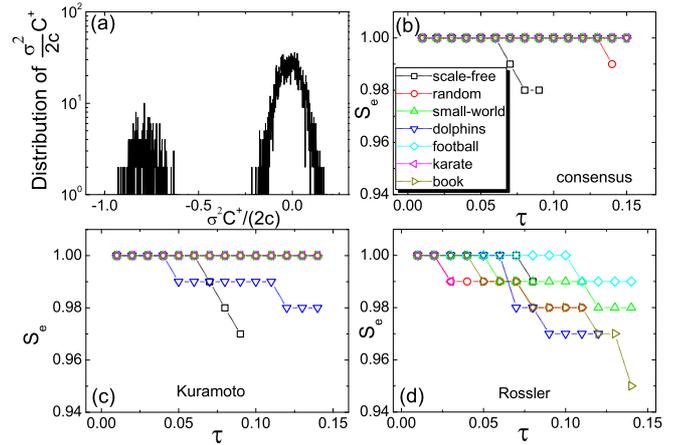,width=\linewidth} \caption{(Color online)
(a) Example of the distribution of the values of elements of the
generalized inverse $\mathbf{{C}}^{\dag}$ of the dynamical
correlation matrix $\mathbf{{C}}$ for consensus dynamics associated
with a scale-free network, where $\tau = 0.05$. The success rate of
prediction of existent links $S_e$ for (b) consensus dynamics, (c)
Kuramoto oscillators and (d) R\"ossler dynamics as a function of
time delay $\tau$ for a number of model and real-world networks:
scale-free networks (scale-free) \cite{BA:1999}, random network
(random) \cite{ER:1959}, small-world network (small-world)
\cite{WS:1998}, dolphin social network (dolphins) \cite{DS}, network
of American football games among colleges (football) \cite{ACF},
friendship network of karate club (karate) \cite{FKC} and network of
political book purchases (book) \cite{PBP}. Other parameters are the
same as in Fig.~\ref{fig:cii}. The success rate of nonexistent links
is higher than 0.99 for all considered cases and thus are not
shown.} \label{fig:fig1}
\end{center}
\end{figure}

Figure~\ref{fig:cii} provides an example of the dependence of
fluctuations $C_{ii}$ on the time delay for three dynamical
processes on both heterogeneous and homogeneous networks. The
results are in good agreement with the theoretical prediction from
Eq.~(\ref{eq:cii}). A non-ignorable deviation from the predicted
fluctuations in noisy R\"ossler dynamics with time delay is
observed, which is mainly due to the simplification of
one-dimensional variable to get Eq.~(\ref{eq:cii}) while the state
variable in R\"ossler dynamics has three dimension. Note that, in
the absence of time delay, the dependence of $C_{ii}$ on the node
degree $k_i$ can be described as a power law \cite{WCHLH:2009,
RWLL:2010}: $C_{ii} \sim k_i^{-1}$, regardless of the dynamics. For
$\tau \ne 0$, the deviation of $C_{ii}$ from the power-law behavior
can then be used to assess preliminarily whether there is a
significant time delay in the underlying networked system: a more
severe deviation suggests a larger value of the time delay.

After calculating the dynamical correlation matrix $\mathbf{{C}}$,
we can infer the details of the network connections through
Eq.~(\ref{eq:infer}) via the generalized inverse of $\mathbf{{C}}$.
Figure~\ref{fig:fig1}(a) shows the distribution of the off-diagonal
elements of $[\sigma^2/(2c)]\mathbf{{C}}^{\dag}$. We observe a
bimodal behavior with two peaks: one centered at a negative value
which corresponds to existent links, and another centered at zero
which indicates non-existent links. Without time delay, the hump in
the distribution for the existent links should be centered at $-1$.
While with time delay, due to the contribution of the term
$c\tau(k_i+k_j)\approx 2c\tau\langle k\rangle$, the center of the
hump will shift toward zero. The larger the time delay, the more
significant the shift will be. For example, as shown in
Fig.~\ref{fig:fig1}(a), the amount of the shift in the negative peak
is $2c\tau\langle k\rangle=0.2$. Considering $c=0.2$ and $\langle
k\rangle=10$ in the example, we obtain $\tau=0.05$, which is exactly
the pre-assumed value of the time delay in the system. To separate
the two humps, a threshold is needed, where all existent links in
the network are identified by the elements in $\mathbf{{C}}^{\dag}$
which lie below the threshold. In particular, the subscript $ij$ of
a picked element below the threshold indicates a link between nodes
$i$ and $j$. We set the threshold at the situs, which corresponds to
the minimum value of the fitted curve between two humps. The
performance of our prediction method can be characterized by the
success rate $S_e$ of the existent links, which is defined as the
ratio of the number of successfully predicted existent links to the
total number of existent ones. As shown in Figs.
\ref{fig:fig1}(b)-(d), our method yields high success rates for
different values of the time delay $\tau$, regardless of the nodal
dynamics and of the network structures.

\begin{figure}
\begin{center}
\epsfig{file=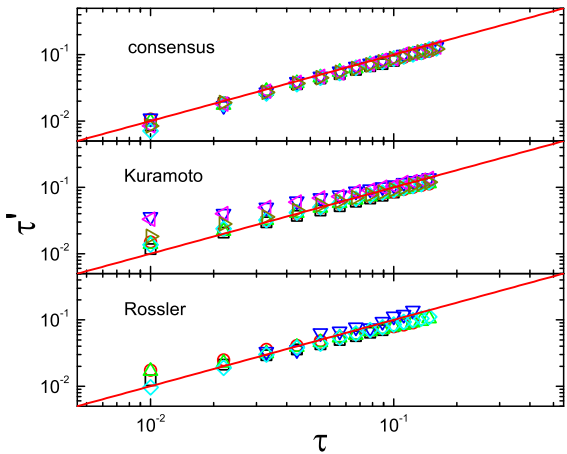,width=0.9\linewidth} \caption{(Color
online) Predicted time delay $\tau'$ from Eq.~(\ref{eq:tau}) versus
the true (pre-assumed) values for the three dynamical processes on a
number of model and real-world networks. The symbols denote the same
networks as in Fig.~\ref{fig:fig1}. The lines are $\tau' = \tau$.
Other parameters are the same as in Fig.~\ref{fig:cii}.}
\label{fig:tau}
\end{center}
\end{figure}

After the network topology $\mathbf{{L}}$ is predicted, we can
estimate the time delay $\tau$ through Eq.~(\ref{eq:tau}). As
demonstrated in Fig.~\ref{fig:tau}, the predicted values of $\tau$
are quite close to the real values for almost all dynamical
processes and network structures considered. The non-ignorable
deviation occurring in Kuramoto dynamics, may be due to its
$\mathrm{sin}$ coupling function which is conditionally approximated
to the linear coupling function used in our theory.

\begin{figure}
\begin{center}
\epsfig{file=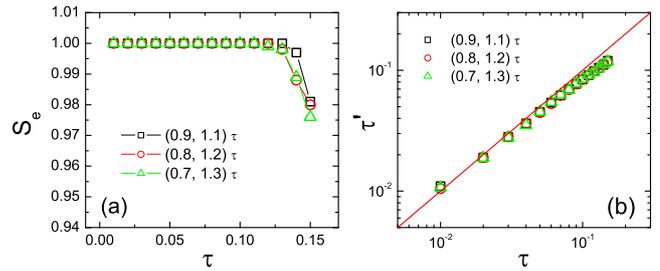,width=\linewidth} \caption{(Color online)
(a) The success rate of prediction of existent links $S_e$ as a
function of the average time delay $\tau$ for different range of
time delays for random consensus networks. (b) The predicted
average time delay $\tau'$ versus the original time delay $\tau$
for different ranges of time delay. The lines are $\tau' = \tau$.
Other parameters are the same as in Fig.~\ref{fig:cii}.}
\label{fig:inhomo_tau}
\end{center}
\end{figure}

We also examine the validity of our method for predicting coupled
network system with inhomogeneous time delay. A random consensus
network associated with random time delays within a certain range
is considered. The success rate $S_e$ as a function of the average
time delay $\tau$ among all pairs of nodes for different ranges of
time delay is shown in Fig.~\ref{fig:inhomo_tau}(a). We see high
success rate if the average time delay is not too large.
Fig.~\ref{fig:inhomo_tau}(b) shows the predicted average time
delay $\tau'$ versus the original time delay $\tau$ for different
ranges of time delay. The predicted $\tau'$ is in good agreement
with $\tau$. These results demonstrate that our approach is
applicable to interacting units associated with inhomogeneous time
delay.

While our theory and the prediction method are based on the system
model Eq. (\ref{eq:main}), a similar theory and method can be
developed for variants of the model. For instance, one can consider
the following system:
\begin{equation} \label{eq:main_2}
\dot{\vec{x}}_i(t)=
\mathbf{F}_i[\vec{x}_i(t)]-c\sum_{j=1}^{N}P_{ij}\big(\mathbf{H}
[\vec{x}_i(t)]-\mathbf{H}[\vec{x}_j(t-\tau)]\big)+ \vec{\eta}_i,
\end{equation}
with the Laplacian matrix in Eq. (\ref{eq:main}) replaced by
$P_{ij}$, the adjacency matrix of the underlying network. The
difference is that, for the dynamics of a given node in Eq.
(\ref{eq:main_2}), the time delay occurs only for the state
information transmitted from its connected neighbors other than the
dynamics of itself. Using similar analytical treatment, one may
arrive at
\begin{equation}
C^\dag_{ij} \approx \frac{2c}{\sigma^2}[(c\tau \langle k\rangle
-1)\mathbf{{P}} -c\tau \mathbf{{P}}^2]_{ij},
\end{equation}
and
\begin{equation}
\tau \approx \left\langle \frac{\left[ \mathbf{{P}} +
\frac{\sigma^2}{2c} \mathbf{{C}}^\dag \right]_{ij}}{c[\langle
k\rangle \mathbf{{P}} - \mathbf{{P}}^2]_{ij}} \right\rangle_{i\neq
j, P_{ij}\neq 0, (\mathbf{{P}}^2)_{ij}\neq 0}.
\end{equation}
The network structures and time delay can again be predicted for
various nodal dynamics and network structures, as we verified
through extensive numerical simulations (data are not shown here).

In summary, we have established a theory to address the inverse
problem for complex networked systems in the presence of time delay
and noise, based solely on measured time series from the network.
Especially, we have obtained a formula relating the generalized
inverse of the dynamical correlation matrix, which can be computed
purely from data, to the structural Laplacian (or adjacency) matrix
and the amount of time delay. Under reasonable approximations the
network topology and the effect of time delay can be separated,
leading to a computationally extremely efficient method for
inferring the network topology and for estimating the time delay.
The validity of the method has been tested numerically using a
variety of combinations of nodal dynamics and network topology,
including a number of real-world network structures. Our method is
completely data driven, and we expect it to be applicable to the
critical network inverse problems in a variety of fields, such as
biomedical and social sciences where complex networked systems are
ubiquitous.

JR thanks Dr. Gang Yan for useful discussions. WXW and YCL are
supported by AFOSR under Grant No. FA9550-10-1-0083.


\begin{thebibliography}{100}
\bibitem{PC:1998}
 \Name{Pecora L. M. \and Carroll T. L.}
 \REVIEW{Phys. Rev. Lett.}{80}{1998}{2109}.
  \Name{Barahona M. \and Pecora L. M.}
 \REVIEW{Phys. Rev. Lett.}{89}{2002}{054101}.


\bibitem{dyn1}
 \Name{Bogu\~n\'a M. \and Pastor-Satorras R.}
 \REVIEW{Phys. Rev. E}{66}{2002}{047104}.


\bibitem{dyn2}
 \Name{Motter A. E., Lai Y.-C. \and Hoppensteadt F. C.}
 \REVIEW{Phys. Rev. Lett.}{91}{2003}{014101}.



\bibitem{dyn3}
 \Name{ Restrepo J. G., Ott E. \and Hunt B. R.}
 \REVIEW{Phys. Rev. Lett.}{97}{2006}{094102}.



\bibitem{dyn4}
 \Name{ Ren J. \and Li B.}
 \REVIEW{Phys. Rev. E}{79}{2009}{051922}.


\bibitem{collins}
 \Name{ Yeung M. K. S., Tegne\'r J. \and Collins J. J.}
 \REVIEW{Proc. Natl. Acad. Sci. USA}{99}{2002}{6163}.

USA {\bf99}, 6163 (2002).

\bibitem{spike}
 \Name{ G\"utig R., Aertsen A. \and Rotter S.}
 \REVIEW{Neural Comput.}{14}{2002}{121};
 \Name{ Pipa G. \and Gr\"un S.}
 \REVIEW{Neurocomputing}{52}{2003}{31};



\bibitem{Lipson}
 \Name{Bongard J. \and Lipson H.}
 \REVIEW{Proc. Natl. Acad. Sci. USA}{104}{2007}{9943}.


\bibitem{Timme:2007}
 \Name{ Timme M.}
 \REVIEW{Phys. Rev. Lett.}{98}{2007}{224101}.


\bibitem{NS:2008}
 \Name{Napoletani D. \and Sauer T. D.}
 \REVIEW{Phys. Rev. E }{77}{2008}{026103}.


\bibitem{WCHLH:2009}
 \Name{ Wang W.-X., Chen Q., Huang L., Lai Y.-C. \and Harrison M. A. F.}
 \REVIEW{Phys. Rev. E}{80}{2009}{016116}.


\bibitem{RWLL:2010}
 \Name{ Ren J., Wang W.-X., Li B. \and Lai Y.-C.}
 \REVIEW{Phys. Rev. Lett.}{104}{2010}{058701}.


\bibitem{Ding:2004}
 \Name{ Dhamala M., Jirsa V. K. \and Ding M.}
 \REVIEW{Phys. Rev. Lett.}{92}{2004}{074104}.



\bibitem{Kanter:2009}
 \Name{ Kinzel W., Englert A., Reents G., Zigzag M. \and Kanter I.}
 \REVIEW{Phys. Rev. E}{79}{2009}{056207}.


\bibitem{Yan:2009}
 \Name{ Yan G., Chen G., Lu J. \and Fu Z.-Q.}
 \REVIEW{Phys. Rev. E}{80}{2009}{056116}.


\bibitem{Scholl:2010}
 \Name{ Flunkert V., Yanchuk S., Dahms T. \and Sch\"oll E.}
 \REVIEW{Phys. Rev. Lett. }{105}{2010}{254101}.


\bibitem{HKS:2010}
 \Name{ Hunt D., Korniss G. \and Szymanski B. K.}
 \REVIEW{Phys. Rev. Lett.}{105}{2010}{068701};
 \Name{ Hunt D., Korniss G. \and Szymanski B. K.}
 \REVIEW{Phys. Rev. Lett.}{105}{2010}{208701};


\bibitem{Gardiner}
\Name{ Gardiner C. W. }
  \Book{Handbook of Stochastic Methods}
   \Publ{Springer, 2ed}
    \Year{1997}.


\bibitem{Horn}
\Name{Horn R. A. \and Johnson C. R.}
  \Book{Topics in Matris Analysis}
   \Publ{Cambridge University Press, Cambridge}
    \Year{1999}.



\bibitem{Saber:2007}
 \Name{ Olfati-Saber R., Fax J. A. \and Murray R. M.}
 \REVIEW{Proc. of the IEEE}{95}{2007}{215}.



\bibitem{Rossler:1976}
 \Name{ R\"ossler O. E.}
 \REVIEW{Phys. Lett. A}{57}{1976}{397}.


\bibitem{Kuramoto:book}
\Name{ Kuramoto Y.}
  \Book{Chemical Oscillations, Waves and Turbulence}
   \Publ{Springer-Verlag, Berlin}
    \Year{1984};
 \Name{ Strogatz S. H.}
 \REVIEW{Physica D}{143}{2000}{1}.


\bibitem{BA:1999}
 \Name{Barab\'asi A.-L. \and Albert R.}
 \REVIEW{Science}{286}{1999}{509}.


\bibitem{ER:1959}
 \Name{ Erd\H{o}s P. \and R\'{e}nyi A.}
 \REVIEW{Publ. Math. (Debrecen)}{6}{1959}{290}.


\bibitem{WS:1998}
 \Name{Watts D. J. \and Strogatz S. H.}
 \REVIEW{Nature (London)}{393}{1998}{440}.


\bibitem{DS}
 \Name{ Lusseau D., Schneider K., Boisseau O. J., Haase P., Slooten E.
\and Dawson S. M.}
 \REVIEW{Behav. Ecol. Sociobiol.}{54}{2003}{396}.


\bibitem{ACF}
 \Name{Girvan M. \and Newman M. E. J.}
 \REVIEW{Proc. Natl. Acad. Sci. USA}{99}{2002}{7821}.


\bibitem{FKC}
 \Name{ Zachary W. W.}
 \REVIEW{J. Anthropol. Res.}{33}{1977}{452}.


\bibitem{PBP}
http://www.orgnet.com/cases.html.

\end{thebibliography}
\end{document}